\documentclass[a4paper,fleqn]{cas-sc}

\usepackage[numbers]{natbib}
\usepackage{rotating}
\usepackage{lineno}
\usepackage{setspace}
\def\tsc#1{\csdef{#1}{\textsc{\lowercase{#1}}\xspace}}
\tsc{WGM}
\tsc{QE}
\tsc{EP}
\tsc{PMS}
\tsc{BEC}
\tsc{DE}

\begin{document}
\let\WriteBookmarks\relax
\def\floatpagepagefraction{1}
\def\textpagefraction{.001}
\shorttitle{Modeling the Resilience of Interdependent Urban Socio-Physical Systems}
\shortauthors{Anonymous et~al.}

\title [mode = title]{Resilience of Interdependent Urban Socio-Physical Systems using Large-Scale Mobility Data: Modeling Recovery Dynamics} 


\author[1]{Takahiro Yabe}[orcid=0000-0001-7511-2910]
\ead{tyabe@purdue.edu}
\credit{Conceptualization of this study, Methodology, Data Analysis, Modeling, and Writing}
\address[1]{Lyles School of Civil Engineering, Purdue University, 550 Stadium Mall Avenue, West Lafayette, Indiana 47907, USA}

\author[1,2]{P. Suresh C. Rao}
\credit{Conceptualization of this study, Methodology, Modeling, and Writing}
\address[2]{Agronomy Department, Purdue University, 915 W State Street, West Lafayette, IN 47907, USA}

\author[1]{Satish V. Ukkusuri}
\cormark[1]
\ead{sukkusur@purdue.edu}
\credit{Conceptualization of this study, Methodology, and Writing}
\cortext[cor1]{Corresponding author}


\begin{abstract}
Cities are complex systems comprised of socioeconomic systems relying on critical services delivered by multiple physical infrastructure networks.
Due to interdependencies between social and physical systems, disruptions caused by natural hazards may cascade across systems, amplifying the impact of disasters. 
Despite the increasing threat posed by climate change and rapid urban growth, how to design interdependencies between social and physical systems to achieve resilient cities have been largely unexplored.
Here, we study the socio-physical interdependencies in urban systems and their effects on disaster recovery and resilience, using large-scale mobility data collected from Puerto Rico during Hurricane Maria. 
We find that as cities grow in scale and expand their centralized infrastructure systems, the recovery efficiency of critical services improves, however, curtails the self-reliance of socio-economic systems during crises.
Results show that maintaining self-reliance among social systems could be key in developing resilient urban socio-physical systems for cities facing rapid urban growth.
\end{abstract}



\begin{keywords}
Resilience \sep Cities \sep Disasters \sep Socio-physical interdependencies \sep Mobility data
\end{keywords}

\maketitle

\doublespacing

\section{Introduction}
Urban communities depend on reliable provision of multiple critical services supplied through infrastructure networks, which are centrally managed by public and private utilities. 
The complexity of such interdependent relationships between social and physical systems embedded within and among urban systems are increasing due to rapid urban growth in many cities around the world \cite{world2009systems}. 
On the other hand, with the rising intensity and frequency of natural hazards globally, there is an increasing need for agencies to enhance the resilience of urban systems to future shocks \cite{unisdr2012disaster}. 
Because of the strong and complex socio-physical coupling, shocks caused by natural hazards may cascade across urban systems, amplifying the disruptions caused by the disaster. 
This poses significant challenges in understanding, modeling and predicting the recovery of cities from future shocks, and identifying operational mechanisms in social and physical networks that enhance the resilience of cities \cite{elmqvist2019sustainability,moglia2018urban}.

The complexity of interdependencies between social and physical systems vary with urban scales. 
Larger cities have bigger social capital to build and manage critical physical infrastructure, and to acquire necessary natural resources (e.g., water, energy, food), and also have access to external technical and financial assistance (subsidy), not readily available to smaller cities. 
Typically, small cities rely on their extant social networks and social capacity inherent in the community. 
During a disaster, due to lack of quick infrastructure recovery, people draw upon their social networks and these are further strengthened \cite{aldrich2015social,waters2017spatial}. 
However, in large cities, due to larger infusions of resources on physical infrastructure, these networks recover quickly \cite{yabe2020regional}. 
The recovery of physical networks is primarily based on resources from a central actor (e.g. government) whereas social networks and the resulting social capital is primarily decentralized. 
As many smaller cities face urban growth, how could these cities, along the rural -- urban growth trajectory, manage the interdependent relationships between social and physical systems for resilient recovery from disasters? 
A key task is to identify the right amount of self-reliance of social systems despite the existence of robust centralized physical infrastructure, that leads to better recovery outcomes for cities based on their size, location and demographics \cite{cutter2014geographies}.
Although studies have explored the role of social and physical networks on urban resilience and recovery in isolation (e.g., social \cite{sadri2018role,mccaughey2018socio,de2019avoiding,cui2020sna}, physical \cite{yang2019physics,mohebbi2020cyber,ouyang2014review,rinaldi2001identifying}, economic \cite{modica2015spatial}), the interdependencies between socio-physical systems and its implications on urban resilience have been largely neglected in existing studies.

\begin{figure*}
\centering
\includegraphics[width=.97\linewidth]{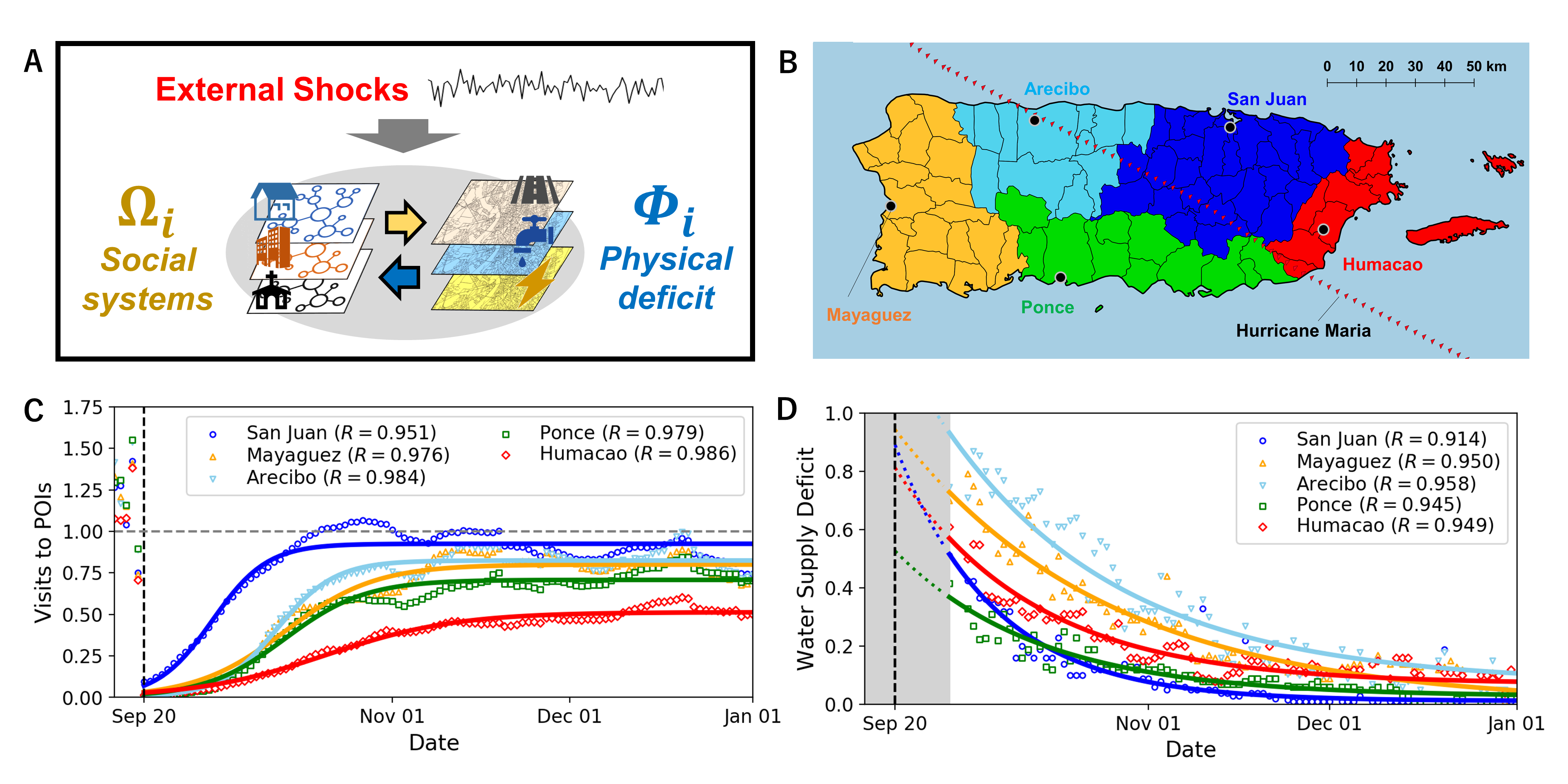}
\caption{Fusion of large-scale data and dynamical model for unraveling socio-physical interdependencies. 
\textbf{A.} Schematic showing the overview of the study. We are interested in how coupled urban socio-physical systems respond to a series of external shocks. Resilience of a CUSPS is quantified using data, system dynamics models, and simulations. \textbf{B.} Puerto Rico was divided into five regions based on the modularity of human mobility flow patterns. \textbf{C.} Recovery of social systems were measured by visits to various places-of-interest (POIs) using mobile phone location data. Disaggregated results are shown in Figure S1. \textbf{D.} Regional differences in recovery of water service deficit after Hurricane Maria in Puerto Rico. Data were available after September 29th 2017, thus observations for initial service deficits between September 20th and September 29th are missing. 
Negative exponential functions ($f(t) \propto \exp{(-\tau t)}$; solid curves) approximate the recovery dynamics well.}
\label{overview}
\end{figure*}

We examine the recovery dynamics of five regions in Puerto Rico island after the devastating damage from the Category 5 Hurricane Maria, to (1) quantitatively assess the temporal trajectories of social and physical recovery, (2) evaluate regional differences in degrees of interdependencies between social and physical systems, and (3) understand the implications of managing such socio-physical interdependencies on urban resilience. 
To achieve these research goals, a data-driven system dynamics modeling approach is applied to estimate the coupled dynamics between social and physical systems during the disaster recovery process (Figure \ref{overview}A).
Empirical data from five regions in Puerto Rico (Figure \ref{overview}B) after Hurricane Maria, including large-scale mobility data collected from mobile phones and recovery data of water infrastructure systems, were used to calibrate and test the dynamics model of coupled socio-physical systems. 
We also extend this model-data analysis to examine the resilience of urban socio-physical systems in the longer time horizon, exposed to a sequence of shocks.
We close with a discussion on strategies on managing the degree of social-physical interdependencies, to design resilient cities as many of them face rapid urban growth in the near future.

\section{Data}
\subsection{Hurricanes Irma and Maria}
Puerto Rico was affected by two significant hurricanes in September 2017. 
Hurricane Irma skirted the northeast area of Puerto Rico as a category 5 hurricane, causing severe flooding on the island. 
Just two weeks later, the tropical cyclone Hurricane Maria made landfall on the southeastern coast of Puerto Rico on September 20, 2017 with wind speeds of 155 miles per hour. 
Heavy rainfall, flooding, storm surge, and high winds caused considerable damage to critical infrastructure in Puerto Rico, causing power outages and water shortages for the entire island for several months \cite{situation}.
Puerto Rico is selected as our case study area and Hurricane Maria as the event in this study, because of its longitudinal impacts on the island and inter-regional heterogeneity in the post-disaster recovery outcomes.

\subsection{Physical Systems Deficit Data}
Data for service deficit of physical infrastructure systems after Hurricane Maria in Puerto Rico were publicly posted on the \texttt{StatusPR} website, which was active until August 2018, which was the following year from the landfall of Hurricane Maria.
The \texttt{StatusPR} website served as a web dashboard that curates physical infrastructure service data from multiple public utility companies in Puerto Rico including gas, water, power, and mobile phone tower connectivity. 
Despite the availability of various physical infrastructure service deficit data, only the water service deficit rates were provided in the regional level; others were all aggregate measures for the entire island.
The water service deficit rates were provided for five regions in the island (metro, north, south, west and east). 
Significant spatial co-location of various physical infrastructure networks with water service networks imply that deficit of water infrastructure systems could serve as an adequate approximation for the aggregated physical infrastructure systems recovery \cite{klinkhamer2017functionally}.
Therefore, in this study, we use water service deficit recovery data to represent the recovery dynamics of physical infrastructure systems.
The recovery dynamics of water service deficit in the five regions are shown in Figure 1D. 

\subsection{Mobile Phone Location Data}
The point of interest (POI) visit dataset was provided by Safegraph Inc (\url{https://www.safegraph.com/}), a company that aggregates anonymized location data collected from smartphone applications to provide insights about physical places.
Safegraph Inc. collects GPS location information from an approximately 10\% sample out of all mobile phones and smartphones in the United States through various apps.
Each GPS data point consists of an anonymized unique user identifier, the longitude, latitude, and timestamp of the observation. 
The longitude and latitude information have high spatial accuracy (typically within 10 meters), thus allows us to analyze visit movements to each POI. 
Users' consent to collect and use their location data were obtained by Safegraph. 

The number of daily visits to each POI were estimated using the mobile phone location data. 
To avoid errors in the estimation, spatial noise in the location data were cleaned by removing jumpy observations where the velocity of the movement was unrealistically high (>150 kilometers per hour). 
The cleaned data points were then spatially clustered to detect ``stay points''using the DBSCAN algorithm \cite{schubert2017dbscan}. 
From the estimated stay points, the visited POI is estimated using a machine learning algorithm that uses various features including: the distance from POI to the stay point, time of day, and likelihood of visits to POI categories (using the North American Industry Classification System code). 
This data-processing procedure produces a time series data of daily visit counts to each POI. 
The POIs in the dataset were categorized into 6 major POI types: education, medical, construction, automobile, grocery, and other department stores. 
Table S2 shows the number of POIs in each region, in each category.

\subsection{Regional Socio-Demographic, -Economic, and Hurricane Damage Data}
Social and economic data of the 78 counties in Puerto Rico were collected from publicly available sources. 
For example, the county population and median income data were retrieved from the American Community Survey (\url{https://www.census.gov/programs-surveys/acs}). 
Housing damage percentages in each county represent the percentage of housing structures that were approved for the Individuals and Households Program of FEMA \cite{FEMA}.

\begin{table}[t]
\small\sf\centering
\caption{Socio-demographic, -economic, and hurricane damage statistics of the five regions in Puerto Rico island.}
\begin{tabular}{crrr}
\toprule
Region & Number of households & Mean Income of households (\$) & Damaged housing due to Hurricane Maria (\%) \\
\midrule
San Juan & 658,457 & 23,823 & 36.2 \\
Mayaguez &  199,355 & 15,614 & 31.4 \\
Arecibo & 143,795 & 16,492  & 38.9  \\
Ponce & 152,153  & 16,505  & 40.8  \\
Humacao & 83,420 & 19,047 & 45.1  \\
\bottomrule
\end{tabular}
\label{stats}
\end{table}

\section{Methods}
\subsection{Network Partition of Regional Mobility Flow}
First, to conduct analysis on the regional heterogeneity in social and economic recovery trajectories, the entire island is partitioned into appropriate geographical regions.
Such geographical regions should reflect the spatial boundaries of daily livelihood and activity patterns of the residents. 
An ideal partitioned sub-network would include a set of counties with a large amount of internal mobility flow among them (e.g. core city and periphery cities) in the same regional cluster, but separate counties that have small amount of mobility flow among them. 
More formally, given a directed bi-directional network of mobility flow patterns (from mobile phone movement data), we attempt to find an optimal way to partition the network into subnetworks so that we achieve high network modularity. 
Various computational algorithms have been proposed to tackle this problem in the context of community detection in networks (or graphs) \cite{fortunato2016community}. 
Among the variety of network partitioning algorithms, the \texttt{infomap} algorithm is shown to be an effective and computationally efficient method \cite{rosvall2009map}, and has been applied to network community detection tasks in various domains, including human mobility modeling \cite{bagrow2012mesoscopic}.
\texttt{infomap} is an information-theoretic graph partitioning method, which uses the flows of random walkers to find groups of dynamically related nodes in directed, weighted network. 
The algorithm was implemented on Python, using the \texttt{igraph} package for implementation (\url{https://igraph.org/r/doc/cluster_infomap.html}). 

\subsection{Socio-Physical Systems Dynamics Model}
Klammler et al. \cite{klammler2018modeling} proposed a unique approach for modeling the resilience dynamics of socio-physical (or ``technological-social'') systems. 
The dynamics of the social and physical systems are characterized by coupled differential equations based on modeling insights from the social and ecological sciences. 
The adaptive capacity of social systems $\Omega(t)$ and service deficit of physical systems $\Phi(t)$ are described using the following differential equations in the original model:
\begin{eqnarray}
    \frac{d\Phi}{dt} &=& (1-\Phi)b- w \Phi \Omega + \xi  \\
    \frac{d\Omega}{dt} &=& (1-c_1\Phi)\Omega(1-\Omega) - r \frac{\Omega^n}{\Omega^n + \beta^n} - c_2 \xi
\end{eqnarray}
where, $b$, $r$, $\beta$ and $n$ are model parameters that characterize the functionality of the systems, $c_1$ and $w$ are parameters that describe the strength of coupling between the social and physical systems, and $\xi$ represent the external shocks (i.e. natural hazards) that affect the system. 
$c_2$ controls how much the social systems are affected by the external shocks.
Each of the equations are composed of three components: replenishment (or improvement), depletion (or degradation), and external shock.
The equation of physical systems dynamics $\Phi$ is characterized by an exponential growth term of degradation (parameterized by $b$), exponential recovery which depends on the social system state and parameter $w$, and an external shock $\xi$. 
The equation of social capacity dynamics $\Omega$ is characterized by a logistic replenishment function term, which depends on the physical deficit state by parameter $c_1$, a Hill type function representing the depletion of social capacity parameterized by $r$, $n$ and $\beta$, and an external shock multiplied by the impact factor $c_2$.
Simulation results using synthetic data showed that even without external shocks with severe intensity, a series of small but repetitive shocks may tip the system over to a stable undesirable state \cite{klammler2018modeling}.  
While this model was limited to the theoretical discussion of the model and lacked empirical validation, \cite{krueger2019resilience} applied the model to the context of water systems, and evaulated the resilience of water systems of various cities around the world. 
The model parameters were assigned for each city based on a capital portfolio approach that quantifies various urban characteristics \cite{krueger2019quantifying}. 

In this study, we modify this model by relaxing the assumption that physical recovery is deterministically dependent on social systems recovery, and to allow the socio-physical system to have no functional coupling.
To adjust the model to meet our objective, we reformulate the model by introducing an additional model parameter $q$, that represents physical system recovery which is independent of social system recovery. 
Moreover, to fit the dynamics model to data from Hurricane Maria, we remove the repetitive shock sequences $\xi$, and represent the shock impact using the initial disruption values of $\Phi$ and $\Omega$, denoted as $\Phi_0$ and $\Omega_0$, respectively. 
To simplify the dynamics of the model, and to obtain better convergence probabilities in the estimation of model parameters, we fix $n=1$. 
Moreover, we study the socio-physical recovery dynamics for each region $i=\{1,2, \dots, I \}$. 
Thus, the full model in this study is as follows:
\begin{eqnarray}
    \frac{d\Phi_i}{dt} &=& (1-\Phi_i)b_i- (w_i \Omega_i + q_i) \Phi_i \\
    \frac{d\Omega_i}{dt} &=& (1-c_i\Phi_i)\Omega_i(1-\Omega_i) - r_i \frac{\Omega_i}{\Omega_i + \beta_i} \\
    && i=1,2,3,...,I
\end{eqnarray}
where, initial conditions are given by $\Phi_i(0) = \Phi^i_0$, and $\Omega_i(0) = \Omega^i_0$, which are also model parameters.
Using this formulation, we are able to estimate and analyze the regional heterogeneity in the system dynamics and model parameters. 
The social dynamics in each region will be analyzed for multiple POI categories $c = \{1,2, \dots, C \}$ as well, to understand the intra-regional heterogeneity in the model parameters. 
The descriptions of the model parameters are summarized in Table S3.
It is important to note that when $c_i = 0$, recovery of social systems are independent of physical deficit states, and when $w_i=0$, physical recovery occurs independently of social system states. 
A system where $c_i = w_i = 0$ denotes a completely decoupled system, which serves as our null hypothesis model. 

\subsection{Estimation of Model Parameters}
To estimate the model parameters $\Theta_i = \{ \Omega^i_0, \Phi^i_0, c_i, r_i , \beta_i, w_i, b_i, q_i \}$, we apply a Hamiltonian Monte Carlo (HMC) sampling approach and obtain the maximum a posteriori (MAP) estimate using empirical observations of social and physical systems collected during the recovery phase after Hurricane Maria. 
Given the discrete time series data of social and physical systems in region $i$, $\omega^i_t$ and $\phi^i_t$, respectively, and the simulated trajectories of social and physical systems of region $i$, $\Omega(\Theta_i)$ and $\Phi(\Theta_i)$, respectively, the likelihood is computed using a Gaussian distribution: 
\begin{eqnarray}
    p(\Omega_t(\Theta_i) ) \sim  \mathcal{N}(\omega_t^i, \sigma_i^2) \\ 
    p(\Phi_t(\Theta_i) ) \sim  \mathcal{N}(\phi_t^i, \sigma_i^2) \\
    c_i, r_i , \beta_i, w_i, b_i, q_i, \sigma_i \sim Cauchy(0,2.5) \\
    \Omega^i_0, \Phi^i_0 \sim Uniform(0,1)
\end{eqnarray}
To allow flexibility, half-Cauchy priors with scale of 2.5 are assigned to the model parameters, including the standard deviation $\sigma$ of the likelihood function. 
The HMC sampler for this MAP estimation is constructed using \texttt{stan}, a Bayesian computational framework.
The sampler drew 5000 samples for each model parameter, and was made sure that the sampler had good mixing by observing that $\hat{R}$ was equal to 1.0.  
The MAP estimate of the model parameters were obtained by taking the mode of the posterior distribution, using Kernel density estimation. 

To evaluate the model fit, the Pearson correlation between the simulated and observed system dynamics was computed.  
Given two time series vectors $x$ and $y$, the Pearson correlation coefficient $\rho_{xy}$ is computed by the following equation:
\begin{equation}
    \rho_{xy} = \frac{cov(x,y)}{\sigma(x)\sigma(y)}
\end{equation}
where, $cov(x,y)$ is the covariance between $x$ and $y$, and $\sigma(x)$ denotes the standard deviation of $x$. 

Furthermore, to interpret the model parameters, multivariate regression was performed on the model parameters using the socio-economic variables listed in Table \ref{stats}.
We investigate which socio-economic variable explains the heterogeneity in the estimated model parameters by selecting the variable with the highest statistical significance in the regression analysis. 
All of the model parameters as well as the socio-economic variables were log transformed prior to the multivariate regression to assure positivity. 

\subsection{Resilience Analysis via Simulation of Urban Socio-Physical Systems}
To measure the resilience of the coupled urban socio-physical systems defined by the systems dynamics model and estimated model parameters, we simulate the longitudinal dynamics under various disaster scenarios. 
Based on the literature, we assume that hurricane occurrence follow a Poisson process with rate of $\lambda$ \cite{rose2012quantifying}. 
Moreover, we assume that the intensity of hurricanes follow an exponential distribution with mean $\alpha$, as assumed in previous modeling literature \cite{klammler2018modeling}. 
According to the Tropical Meteorology Project at Colorado State University and the GeoGraphics Laboratory at Bridgewater State University, Puerto Rico is predicted to have a 8\% probability of 1 or more major hurricanes tracking within 50 miles of the island \cite{hurricanepred}. 
We convert this value to parameters $\lambda$ and $\alpha$ using the following logic.  
The cumulative probability function of an exponential distribution with mean $\alpha$ is given by $F(x;\alpha) = 1-e^{\alpha x}$ when $x>0$. 
Thus, the probability of the shock exceeding $x$ is given by $p=1-F(x;\alpha) = e^{-\alpha x}$. 
Thus, if we assume that a major hurricane has an intensity of 1, in order to have probability $p$, we set $\alpha = - \log p$. 
We assume that hurricanes only occur during the hurricane season (June 1st $\sim$ November 30th), and 500 simulations (each simulating over a 10 year time horizon) were run for each region.  

To investigate the impacts of various policy levers on improving resilience, we test the resilience of urban socio-physical systems under different coupling model parameters ($c,w$). 
More specifically, regional model parameters on efficiency $w$ and dependence $c$ are varied within a range of parameters ($0\leq w\leq 0.1$, $0\leq c \leq 1.5$), while keeping all of the other model parameters (i.e., $r,\beta,q,b$) the same for each region. 
In each simulation run, the collapse time (timestep when $\Omega=0$) was recorded to represent the resilience (= ability to build back after shocks) of the region.

\section{Results}
\subsection{Empirical Observations of Socio-Physical Recovery}
Figure \ref{overview}C shows the normalized visits to various points-of-interest (POIs) in the five regions observed from mobile phone location data, which increasingly have been used to understand urban dynamics \cite{gonzalez2008understanding,blondel2015survey}. 
Several important observations can be made from the time series plotted in Figure \ref{overview}C. 
First, we observe significant differences in the speed of recovery across the regions after Hurricane Maria. 
San Juan region experienced the quickest recovery (in all of the POI types as shown in Figure S1), followed by the Arecibo, Ponce, Mayaguez, and Humacao regions.
Socio-demographic, -economic, and hurricane damage characteristics of the five regions are shown in Table \ref{stats}.
The recovery trajectories, despite the differences in rates, can be well approximated using a logistic curve starting from around zero (complete failure), and asymptotically converging to one (full recovery) over the long term, which corresponds to the pre-hurricane chronic baseline.
We also observe significant increase in visits in all regions just before the landfall of Hurricanes Irma and Maria. 
This indicates substantial preparation activities of the residents before the hurricane (e.g. shopping for grocery and goods), which was also identified in previous studies \cite{yin2014agent}.
Decreases in visits just before Hurricane Irma are much less compared to Hurricane Maria, reflecting the difference in the severity of the hurricanes. 
Further analysis of the disaster impact differences in the region and category of the POIs has been studied in a past study using the same dataset \cite{yabe2020quantifying}.

Figure \ref{overview}D shows the time series data of water service deficit rates in the five regions after Hurricane Maria \cite{statuspr}. 
Negative exponential functions ($f(t) \propto \exp{(-\tau t)}$), shown in solid curves, are shown to fit the recovery of physical service deficit well with Pearson correlation higher than $R = 0.9$. 
This trend agrees with previous observations from other disasters including Hurricane Irma in Florida, Tohoku Tsunami in Japan, and Kumamoto Earthquake in Kyushu, Japan \cite{yaberecovery}.
The recovery rate coefficient $\tau$ is significantly different among the five regions, with San Juan being the quickest and Mayaguez having the slowest recovery.  
The initial value of water supply deficit, which varies among regions but are all below 1, is indicative of estimated initial deficit that was caused by Hurricane Irma.  

\begin{figure*}
\centering
\includegraphics[width=\textwidth]{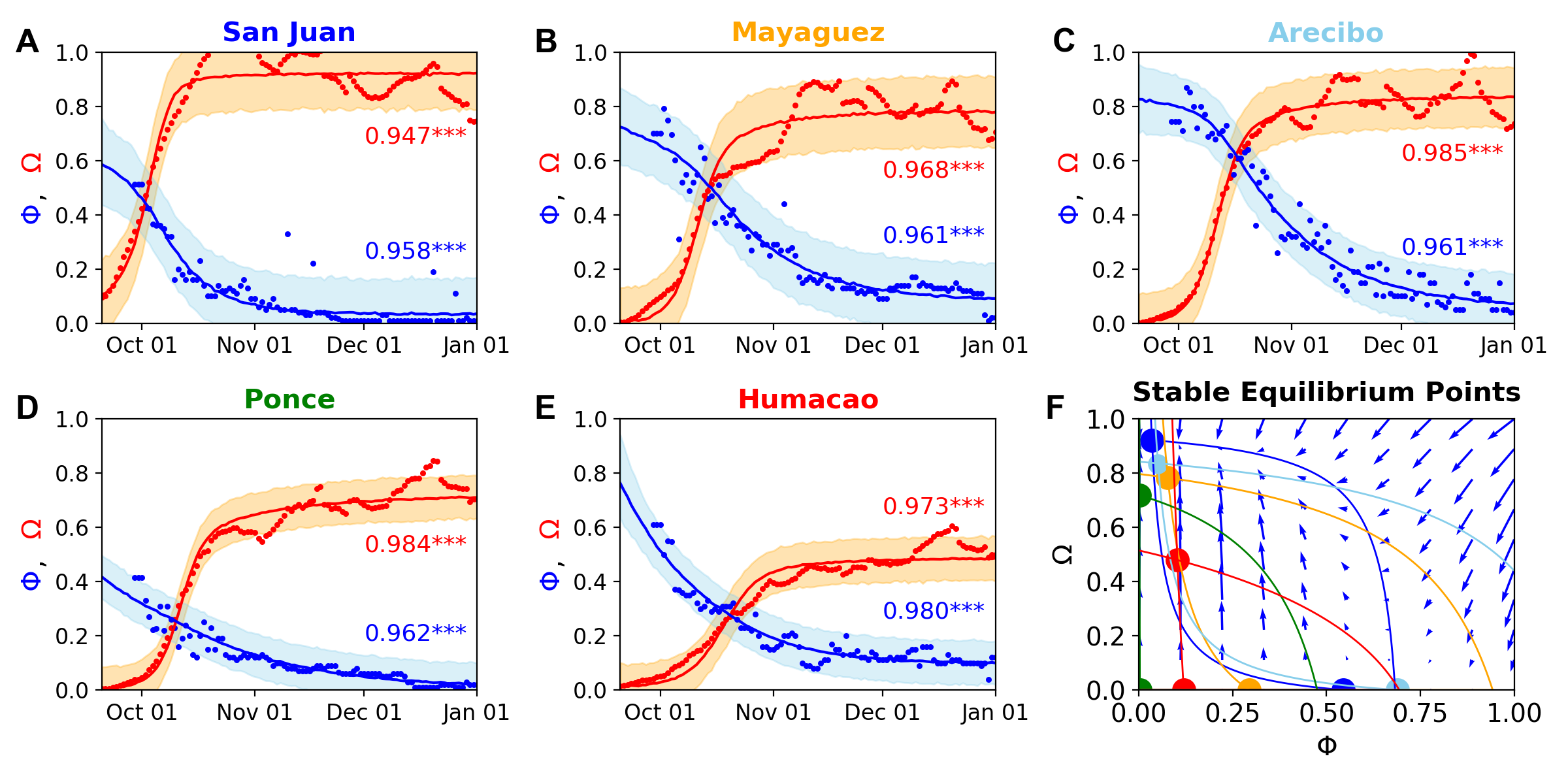}
\caption{Estimation of regional socio-physical recovery dynamics. 
\textbf{A-E. }Observed (dotted) and estimated (solid curves) social (orange) and physical (skyblue) recovery dynamics in each region, using the calibrated socio-physical system dynamics model. The shaded ranges around the simulated dynamics show the 95\% Bayesian credible interval. 
In addition to the strong and significant correlation shown in Table S4, the plots qualitatively show that the system dynamics model is able to replicate the observed dynamics. \textbf{F.} Equilibrium analysis of the five regions. Each region has 2 stable equilibrium points (color filled circles: desirable (high $\Omega$, low $\Phi$) and undesirable ($\Omega=0$)). Direction field for San Juan are shown (arrows).}
\label{fitting}
\end{figure*}

These empirical observations on the dynamic states of social and physical systems are useful in identifying the disparities across the five regions in disaster recovery \cite{cutter2016urban}. 
Although such observations could be informative for monitoring system states, analyzing them in isolation neglects the functional interdependent relationships that exist between the urban social and physical systems. 
Examples of functional interdependencies include: dependence of water networks' (physical systems) recovery on local and federal agencies (social systems), operation of local businesses (social systems) depending on power-grid infrastructure (physical systems), and communities depending on recovery of critical services.
In order to quantitatively capture the functional coupling between the social and physical systems, the empirical observations are integrated with a dynamic model of coupled socio-physical systems.

\subsection{Inference of Socio-Physical interdependencies}
The coupled dynamics model of socio-physical systems, originally proposed by Klammler et al. \cite{klammler2018modeling}, is composed of two coupled differential equations (\textbf{Methods}). 
This model is applied due to its high similarity with empirical observations, especially the logistic recovery curve of social systems (Figure \ref{overview}C) and exponential recovery curve of physical systems (Figure \ref{overview}D). 
In this study, we revised the model by including an additional parameter $q$ in Equation (1) to relax the assumption that the recovery of physical systems deterministically depend on social systems recovery; the second term $-(w\Omega + q)\Phi$ is the general form of the original $-w\Phi\Omega$ term.  
Among the model parameters $\Theta = \{\Omega_0, \Phi_0, c, r , \beta, w, b, q \}$, $c$ and $w$ dictate the coupling strength among the socio-physical systems, and completely decoupled urban systems can be characterized with $c=w=0$. 
Descriptions of the model parameters are listed in Table S3.
The model parameters were calibrated to the data in the five regions using Hamiltonian Monte Carlo (HMC) sampling methodology and maximum a posteriori (MAP) estimation (\textbf{Methods}). 
Figure \ref{fitting} A-E show the observed social and physical recovery dynamics (in dotted lines) against the calibrated social and physical simulation dynamics (in solid lines), colored in orange and blue, respectively, for each region in Puerto Rico.
The plots show the high reproducibility of the socio-physical recovery dynamics model, with high Pearson correlation (all higher than $R=0.9$) between the data and the model. 
The estimated model parameter values (mean and 95\% credible intervals) for the five regions as well as for different point-of-interest categories are shown in Table S4. 
The generalizability of the socio-physical dynamics model, was evaluated via testing on recovery data of different point-of-interest (POI) categories (Figure S3, Table S2), including education, medical, construction, automotive, grocery, and other stores (Figure S1). 
The fitting results (Figure S4, Table S4) show that the socio-physical dynamics model is capable of evaluating industry level recovery.
The stability of the socio-physical systems in the five regions was analyzed. 
Figure \ref{fitting}F shows the phase plane of the socio-physical systems. 
The stable equilibrium points (filled circles), null-clines (solid lines), and direction field (arrows; only for San Juan is shown) are shown in the diagram. 
The phase planes of each region are shown in Figure S5 in detail. 
The socio-physical systems in the five regions each have 2 stable equilibrium points, 1 being a desirable equilibrium (high social recovery $\Omega$, low physical service deficit $\Phi$), and 1 being an undesirable equilibrium ($\Omega=0$). 
The desirable equilibrium state of Humacao (red) in particular, is $\Omega \sim 0.5$ and $\Phi \sim 0.13$, which highlights the long-term chronic deficit compared to other regions.

\begin{figure*}
\centering
\includegraphics[width=.97\textwidth]{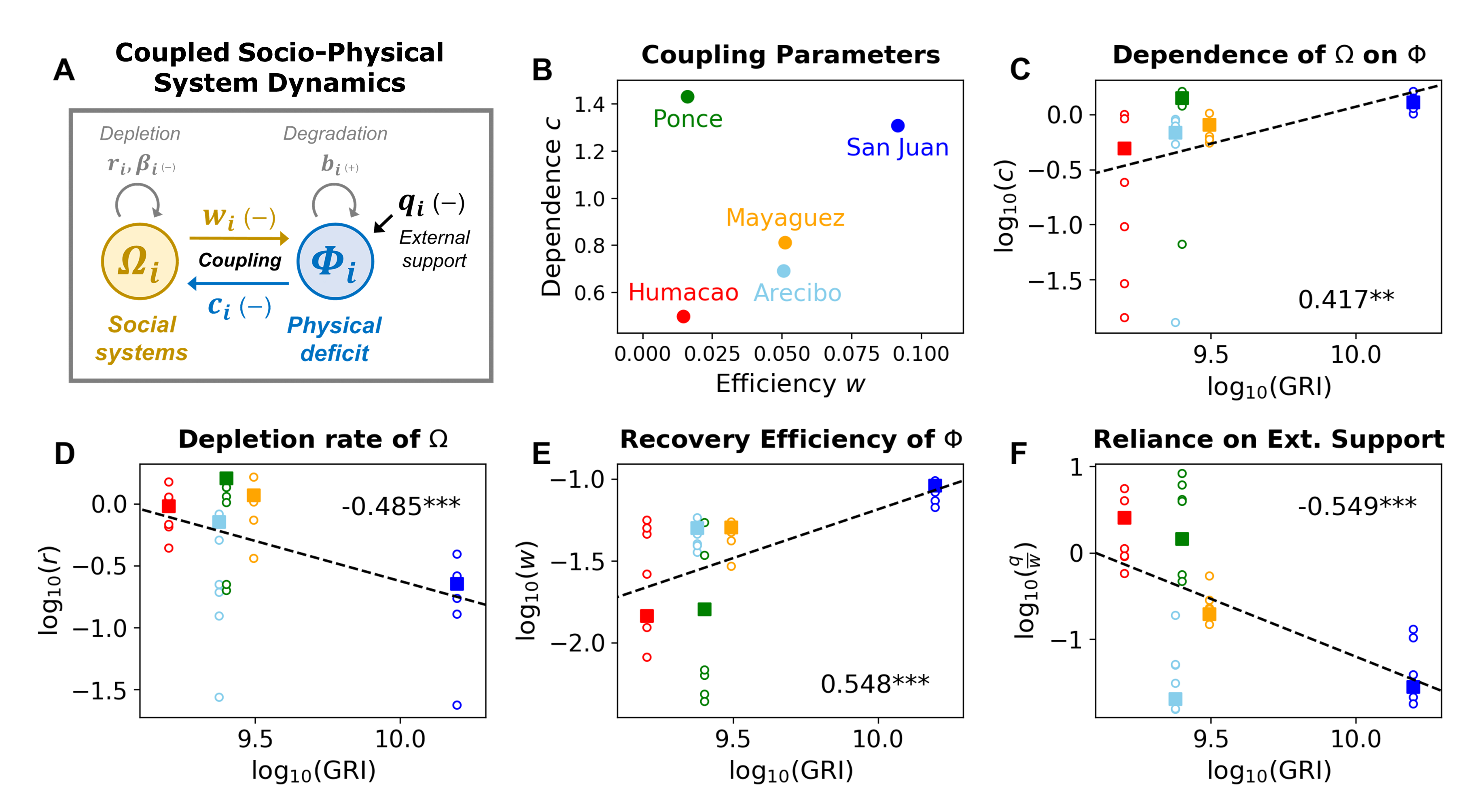}
\caption{Regional differences in system dynamics model parameters. 
\textbf{A. } Setup of the coupled socio-physical system dynamics model and the model parameters.
\textbf{B. } Estimated coupling parameter values for the five regions suggest a trade-off relationship between efficiency in recovery ($w$), and dependence on physical infrastructure systems ($c$).  
\textbf{C-F.} Correlation between the estimated model parameters ($c,r,w,\frac{q}{w}$) and the gross regional income (GRI) in log-log plot. 
Colors correspond to the five regions in Figure S2, and the dotted black line shows the linear regression of the logged variables. 
The Pearson correlation coefficient is shown in the bottom corner of each plot, with stars indicating its statistical significance (***: $p<0.01$, **:$p<0.05$, *:$p<0.1$). 
Significant correlation was observed, indicating that regions experience different recovery dynamics based on their GRI.
}
\label{corrs}
\end{figure*}

\subsection{Urban Scale and Socio-Physical interdependencies}
\begin{figure*}
\centering
\includegraphics[width=\linewidth]{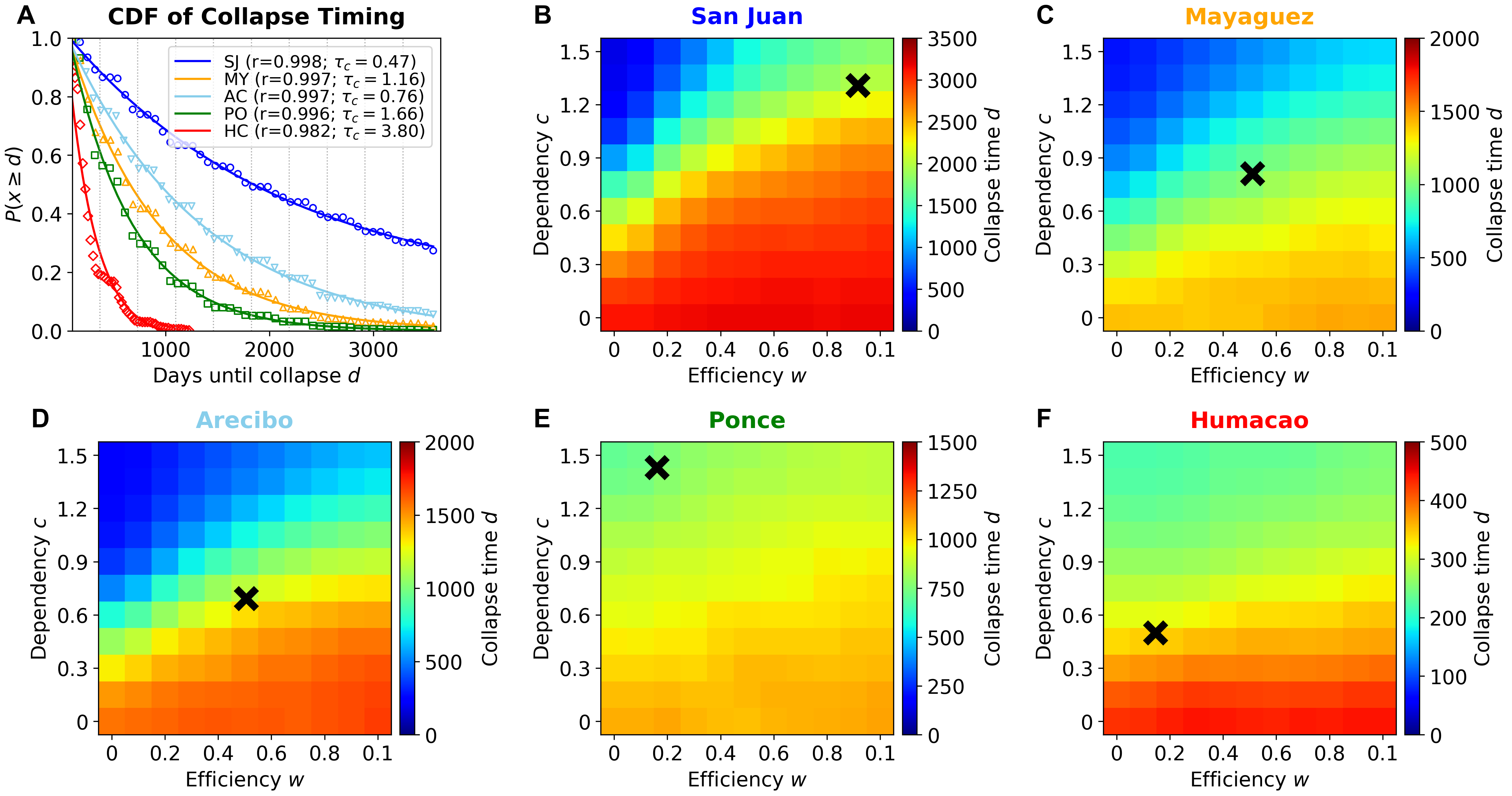}
\caption{Resilience implications of socio-physical interdependencies. \textbf{A.} Cumulative density functions of collapse timing across the shock sequence simulations in the five regions. San Juan and Humacao are shown to be the most and least resilient, respectively. \textbf{B-F.} Resilience implications of reforming the interdependencies (efficiency $w$ and dependence $c$) between the social and physical systems. In all regions, increasing efficiency and lowering the dependence on physical systems benefit regional resilience, although the marginal effects vary across regions. Note that the color bar scales vary across panels.}
\label{policysim}
\end{figure*}

The coupled urban socio-physical model and the estimated parameters allow us to further understand why recovery patterns were so heterogeneous across regions in Puerto Rico. 
Among the model parameters shown in Figure \ref{corrs}A, $c$ and $w$ govern the strength of interdependencies that exist across the social and physical systems (i.e. ``coupling parameters''). 
Parameter $c$ (``\textit{\textbf{dependence}}'') controls to what extent physical service deficit slows down the recovery of social systems. 
Larger \textit{dependence} indicates that social systems are highly dependent on the recovery of physical systems, lacking self-reliance.
Parameter $w$ (``\textit{\textbf{efficiency}}'') controls how efficiently the social systems are able to restore damaged physical systems. 
Larger \textit{efficiency} indicates higher recovery capacity of social systems.
Figure \ref{corrs}B shows the parameters $c,w$ of the five regions. 
The estimated parameters suggest a trade-off relationship between efficiency $w$ and dependence $c$, where San Juan (blue) (most populated, higher average income) have high recovery efficiency but the social systems have high dependence on the physical systems, and on the other hand, Humacao (red) (less populated) has less recovery efficiency but are more self-reliant. 
Arecibo (skyblue) and Mayaguez (orange) (intermediate population density) are placed in between San Juan and Arecibo regions. 
Although Ponce (green) had less recovery efficiency, its social systems had high dependence on physical systems. 
The implications of these coupling parameters on the resilience of the regions are further investigated and discussed in the following section (Figure \ref{policysim}). 

To further interpret the estimated model parameters, multivariate analysis was conducted on the model parameters of the POI types in the five regions $\Theta$ using socio-economic variables, including the total number of housing damage rates, mean income, and gross regional income (GRI) of each region, shown in Table \ref{stats}. 
As a result, it was found that $\log_{10}$(GRI) had statistically significant correlation with model parameters ($c,r,w$), 
Panels C-F in Figure \ref{corrs} show the significant correlation (weak scaling) between the model parameters $c,r,w,\frac{q}{w}$ and GRI. 
The color-filled square plots show the parameters for the aggregated regional dynamics, and open circles show the estimated parameters for each POI category. 
The strong positive correlation between GRI and $c$, and between GNI and $w$ indicate that regions with more population and income are more efficient in recovery, but also more dependent on physical systems. 
This shows that the bi-directional dependencies between social and physical systems grow stronger as cities become larger and wealthier. 
This supports our hypothesis that larger urban systems are embedded within more complex interdependencies between social and physical systems.
This also agrees with the analysis in Padowski et al. \cite{padowski2016overcoming} on water systems in several cities in US and Africa, which showed how managers of larger cities need to overcome the lack of local resources by constructing more complex social and physical frameworks.  
Such wealthier regions also are affected by lower depletion rates of social systems ($r$), which suggest better maintenance capacities.
Moreover, these regions require less external support for physical recovery, in relative terms with respect to their internal recovery efficiency ($\frac{q}{w}$).

\subsection{Resilience Implications of Socio-Physical interdependencies}
So far, we have unraveled the strengths of interdependencies between the social and physical systems during recovery from a shock using a large-scale data-driven system dynamics modeling approach. 
However, it is not clear how the interdependent relationships would affect the resilience of the socio-physical systems in the longer time horizon, when these systems are exposed to multiple shocks. 
To assess the resilience of the calibrated socio-physical systems, we performed Monte Carlo simulations to examine the response of the system to a sequence of stochastic shocks (\textbf{Methods}). 
The sequence of shocks are randomly generated based on hurricane intensity predictions provided by studies in the climate modeling literature.
We assume that hurricane occurrence follow a Poisson process \cite{rose2012quantifying}, and their intensities follow an exponential distribution \cite{klammler2018modeling}. 
We calibrate the hurricane model parameters using the Tropical Meteorology Project at Colorado State University and the GeoGraphics Laboratory at Bridgewater State University, which predicts Puerto Rico to have a 8\% probability of 1 or more major hurricanes tracking within 50 miles of the island \cite{hurricanepred}. 
We assume that hurricanes only occur during the hurricane season (June 1st $\sim$ November 30th). 
500 simulations (each simulating over a 10 year time horizon) were run for each region.  In each simulation run, the collapse time (timestep when $\Omega=0$) was recorded to represent the resilience (= ability to build back after shocks) of the region (\textbf{Methods}). 
Figure \ref{policysim}A shows the cumulative probability densities of collapse timings in each region. 
San Juan is the most resilient, while Humacao has a significantly low probability of avoiding collapse beyond 1000 days. 

Using this simulation framework, we investigate whether by reforming the strengths of interdependencies between the social and physical systems (model parameters $c,w$), the resilience of these regions in Puerto Rico could be improved.
To address this question, regional model parameters on efficiency $w$ and dependence $c$ are varied within a range of parameters ($0\leq w\leq 0.1$, $0\leq c \leq 1.5$), while keeping all of the other model parameters (i.e., $r,\beta,q,b$) the same for each region. 
Similar to the aforementioned framework, the impacts of reforming coupling parameters on the regional resilience (this time, using mean collapse time) are simulated.
Panels B-F in Figure \ref{policysim} show the resilience of the five regions under different strengths of socio-physical interdependencies. 
The black crosses indicate the current situation of socio-physical interdependencies in the regions. 
Warmer colors indicate longer collapse times (more resilient), while colder colors indicate quicker collapse times (less resilient). 
In all of the regions, we observe that improving recovery efficiency $w$ and lowering the dependence on physical systems $c$ lead to resilient urban systems. 
However, note that the marginal improvements of the two levers vary for different regions.
While San Juan, Mayaguez, and Arecibo could gain more resilience by shifting in both directions (right-wards and down-wards), Humacao may yield little marginal gain by strengthening its physical efficiency (right-wards). 
Rather, Humacao should further decrease its dependence on physical infrastructure and strive towards a self-sustainable and decentralized system to improve its resilience to future shocks. 
Reforming the socio-physical interdependencies is one policy lever that can be implemented to enhance resilience. 
Other policies include improving the robustness of physical infrastructure, which can be also simulated with our model by applying a buffer that reduces the magnitude of shocks when they are below a certain threshold (\textbf{Methods}).
Improving the robustness of infrastructure is also shown to be effective in improving the resilience of these regions (Figure S7), which agrees with current practices (e.g., building structures such as sea walls, water drainage systems). 
To enhance the resilience of urban systems to future shocks, it is crucial to not only focus on the structural improvements but also to maintain self-reliance of social systems, especially in urban areas.

\section*{Discussion}
In many OECD countries, reliable critical services (e.g., water, sewage, power, transport) are available on demand, provided by robust connectivity to efficiently (and often centrally) managed critical infrastructure systems. 
On the other hand, cities in less developed regions and countries have less reliable provision of such services, thus, citizens often utilize a wide array of adaptive strategies to cope and to overcome service deficits of critical infrastructure systems \cite{krueger2019resilience,klingel2012technical}. 
For example, household interviews have found that citizens in the Humacao region, which were most heavily affected by Hurricane Maria, supported eachother in the absence of critical physical infrastructure services (e.g., ``... Her neighbors and the community bakery allow her to store cold food in their refrigerators.'') \cite{humacao}. 
In such regions, households use reserves (e.g., pre-positioning critical supplies or boarding homes) or seek assistance from alternate providers (sharing generators; purchasing bottled water) \cite{krueger2019quantifying}.
As more cities face urbanization and robust physical infrastructure are built, the dependence of social entities (e.g., households, businesses) on such physical systems will increase. 
As suggested in San Juan's case in this study, high robustness and efficiency of physical infrastructure could lead to changes in people's behavioral patterns, putting higher dependence on physical infrastructure, similar to the citizens in OECD countries. 
The exception seems to be Ponce, which had both high dependence and low efficiency (Figure \ref{corrs}A).
This was because Ponce was not in the direct path of the hurricane, and the region was given more financial support from federal agencies such as FEMA \cite{femaassistance}.
However, with increasing frequency and intensity of climate related hazards, recent disasters have shown the risk of over-reliance on physical systems, as no engineered system is fail-proof \cite{yu2020toward}.
The simulation results obtained in this study reinforce this point, based on empirical data and modeling, that increasing dependency on physical systems decreases the resilience of communities. 
Therefore, the proposition of this study -- the importance of maintaining self-reliance of social systems -- will become key as cities simultaneously face rapid urbanization and climate change. 

Enhancing community-level  resilience requires trade offs at multiple spatial and temporal scales \cite{chelleri2015resilience,krueger2020balancing} between security at household scales to sustainability at larger scales. Similarly, trade offs need to be made between increasing robustness of the physical (engineered) infrastructures or decentralizing their management. Adaptive capacity required to cope with chronic disturbances and major shocks requires balancing diverse "capitals" (see \cite{krueger2019quantifying}), which are unequally distributed within and among urban communities in a region. Thus, optimizing interdependencies and resource flows among affected communities is another path to enhancing regional resilience.  

Another challenge in building community resilience is that persistent inequalities exist in adaptive capacity within and between urban communities \cite{krueger2019quantifying}. 
Poorest and marginalized communities suffer the most during disasters, and lacking adaptive capacity or access to external subsidies, recover the last. Thus, regional community resilience must ensure equitable access to reliable critical services \cite{logan2020reframing}.

The insights presented in this study could be applied in policy making to provide more resilient urban systems and favorable recovery outcomes after disasters \cite{grafton2019realizing}. 
Together with detailed estimations of costs to implement various policy levers (e.g., decreasing household dependence by installing power generators in rural areas, improving connectivity of social networks enabling more social capital in urban areas), policy makers will be able to perform cost-benefit analysis for enhancing regional resilience. As shown in Figure S7, we can simulate different hazard scenarios. 
As discussed in this paper, the extent of self-reliance should be carefully weighed based on the type of community (urban versus rural), the starting points of the physical and social networks, the socio-demographics and the local institutional rules that govern the recovery.
The marginal benefit from an improvement in social networks versus physical networks vary across urban and rural areas. 
For instance in Humacao, we observe that if the physical networks are not improved from their current situation, improvements in social networks can only result in small improvements in the collapse time (Figure \ref{policysim}F). 
A key finding is that rural communities need to have a base level of physical network efficiency for them to be resilient. 

By taking into account deep uncertainties in future climate conditions, macroeconomic trends, and demographic changes, a robust decision making framework can be applied to the coupled socio-physical dynamics model to make robust policy decisions \cite{hallegatte2012investment}. 
Although the model was evaluated on data collected from Puerto Rico after Hurricane Maria, the model is generalizable to any type of disaster in any given region or city. 
Recently, data collected from mobile devices have been increasingly used for post-disaster assessment of population dynamics, which can be used to capture the recovery of social systems \cite{wilson2016rapid,yaberecovery,lu2012predictability}. 
Applying the coupled socio-physical dynamics model to other regions of various characteristics could generate insights on its resilience in future climate scenarios.

Several limitations in the proposed approach and results open up various research opportunities. 
First, investigation at a finer-spatial resolution, for example on a county or census tract scale, could provide more detailed results and estimations that could be utilized by decision makers in municipal governments. 
However, down-scaling the analysis could bias the estimations with data sparsity. 
A more robust model parameter estimation method could be a topic worthy of investigation for future studies. 
A finer grained analysis could allow a more detail analysis on socio-demographic inequalities and equitable resilience to disasters \cite{hewawasam2020equitable,yabe2020effects,verschuur2020prioritising}. 
On the other hand, previous studies have focused on larger cities, often with several million households \cite{krueger2019resilience}. 
Extending this data-model approach towards both microscopic and macroscopic directions would be an interesting next step. 
Moreover, the resilience simulations assumed that changing the coupling parameters do not affect the other model parameters of social and physical systems. 
Further investigation relaxing this assumption is needed to obtain a full picture of the resilience implications.  
Second, the data on the recovery of physical systems on the regional scale used to calibrate the model was limited to just water service deficit, due to the lack of available data in Puerto Rico after Hurricane Maria for the other types of infrastructure systems. 
Collection of data for other physical infrastructure including power, gas, transportation systems, more specifically a regionally disaggregated time series data on the service deficit could allow us to extend the analysis to a multi-layer physical network. 
Third, a scaling relationship between the estimated model parameters and regional gross income can be found in Figure \ref{corrs}. 
Bettencourt et al. \cite{bettencourt2007growth} have shown that the generalizability of scaling patterns by testing various urban metrics including total wages, total electricity consumption, and total road length. 
Similarly, the results should be examined further by using datasets of recovery after other events in different cities, regions, and across different disaster types.

\section{Conclusion}
The complexity of cities are increasing due to rapid urbanization around the world. 
Such interdependencies between social and physical systems could amplify the impact of disruptions caused by natural hazards, posing a threat to cities within an impacted region as we face climate change. 
In this study, we proposed a data-driven modeling framework to infer the socio-physical interdependencies in urban systems and their effects on regional-scale disaster recovery and resilience.
Large-scale mobility data collected from mobile phone users in Puerto Rico during Hurricane Maria were used to calibrate the model across five regions within the island. 
Estimation results indicated that as cities grow in scale and expand their centralized infrastructure systems, the recovery efficiency of critical services improves, however, curtails the self-reliance of socio-economic systems during crises, posing a trade-off in urban management.
Further longitudinal simulation results using hypothetical future climate scenarios showed that maintaining self-reliance among social systems could be key in developing resilient urban socio-physical systems for cities facing rapid urban growth.
Economic expansion and population growth in larger cities increase community demands for critical services based on resources drawn from increasing regional scales. 
Migration from smaller cities to larger cities adversely impacts the socioeconomic well being of smaller communities, while overwhelming the existing critical infrastructure, a problem most evident in growth of informal settlements in mega-cities in Asia, Africa, and South America.  
Thus, evaluating and managing community resilience at regional scales is of increasing importance.
These results encourage a paradigm shift in urban planning -- to carefully assess the complex interdependencies between social and physical systems -- to improve regional resilience of urban systems to future shocks.
\printcredits

\bibliographystyle{cas-model2-names}

\bibliography{cas-refs}

\end{document}